\documentclass[]{aastex631} 

\usepackage[utf8]{inputenc}

\shorttitle{Extinction Curves versus Extinction Laws}
\shortauthors{F. Zagury}

\begin{document}

\title{Extinction Curves, Extinction Laws, and the Failure of Interstellar Dust Models }
\author[0000-0002-0786-7307]{Fr\'ed\'eric Zagury}
\affiliation{Fondation Louis de Broglie \\
23 rue Marsoulan \\
75012 Paris, France}

\nocollaboration{1}

\begin{abstract}
The interpretation of ultraviolet Galactic interstellar extinction curves is obscured today by accumulated assumptions, such as a purported link between the 2200~\AA\  bump and metallicity, that are not firmly supported by observations. 
In this paper I define extinction curves as the ratio $F_\star/F_0$ of the near-infrared-to-ultraviolet spectrum of a reddened star to that of the same star without intervening material, rather than in terms of a magnitude difference, and revisit their observed properties.
Special attention is given to the connection that Galactic extinction curves with a 2200~\AA\ bump retain with the ultraviolet extrapolation of the exponential extinction law defined by their near-infrared–to–optical segment.
This connection leads to the classification of all extinction curves into three types.
A graphical representation of these types together with their underlying exponential extinction laws demonstrates that interstellar extinction curves can be interpreted in two ways.
Either they result from the mixing of distinct extinction laws associated with different particles, as traditionally assumed, or Galactic ultraviolet curves with a bump are not extinction laws proper but instead deviate from a universal exponential extinction law owing to an additional contribution from coherently forward-scattered starlight. 
Given the observational constraints on the interpretation of extinction curves, such as their dependence on just two parameters, and the fact that bump-like extinction curves are barely observed outside the Galaxy, the latter interpretation emerges as the only logically consistent one. 
\end{abstract}
\section{Introduction: Observed Extinction Curves and Physical Extinction Laws} \label{intro}
Interstellar extinction was long assumed to be uniform throughout the Galaxy, with an optical depth $\tau_\lambda$ proportional to the reddening $E(B-V)$ and varying  as the inverse wavelength $1/\lambda$ over most of the optical spectrum \citep{hall37, greenstein37, stebbins43,nandy64}. 
However, observations were  confined to the visible and infrared domains, and the advent of  space-based ultraviolet spectroscopy in the 1960s and early 1970s overturned this picture. 
Toward Galactic O- and B-type stars, the ultraviolet part of the magnitude difference between a reddened star and its unreddened counterpart, expected to be $1.08\tau_\lambda$, was found to vary strongly from one direction to another, even over very small angular separations and for the same $E(B-V)$ \citep{bless72}. 
In addition, while the infrared–optical interstellar optical depth follows a $1/\lambda^p$  law with index $p$ close to 1, as expected from a dust size-distribution behavior similar to that observed in planetary or terrestrial atmospheres \citep{rv2}, ultraviolet extinction is characterized by a continuum lacking a simple $1/\lambda$ dependence and by a strong absorption feature near 2200 Å, neither of which corresponds to any known natural process or laboratory-reproducible material. 
Outside the Galaxy, from nearby galaxies to distant quasars, this distinctive Galactic ultraviolet behavior becomes  a minority case; instead, extragalactic extinction typically lacks a bump and extends the optical  extinction law into the ultraviolet (see Section~\ref{ecdis}).

The Galactic extinction pattern has  shaped the study of interstellar extinction. 
Together with the rapid increase in the number of diffuse interstellar bands (DIBs) in the visible spectrum and the discovery of broad unidentified infrared emission bands (UIBs), also during the 1960s–1970s, Galactic ultraviolet extinction motivated the development of interstellar dust models based on the hypothetical existence  of complex molecules such as polycyclic aromatic hydrocarbons (PAHs) in interstellar space.

Over the past decades, these models have multiplied, but none has proved compelling.
No agreement has been reached on the exact nature of the particles responsible for either the bump or the ultraviolet continuum. 
None of the proposed molecular carriers has been specifically identified, either observationally or in laboratory experiments, and Galactic-like extinction has never been reproduced experimentally. 
In addition, existing interstellar dust models possess intrinsic limitations, including an excessive number of free parameters that cast doubt on their reliability \citep{param}.
Yet, these models now constitute the standard framework for analyzing interstellar extinction observations \citep{habart24, hora26}, raising legitimate doubts about the robustness of many published interpretations.

This paper returns to the original observational evidence on interstellar extinction, from the infrared to the ultraviolet, and examines the conclusions that can logically be drawn from it.
The basic data are extinction curves, which are  defined as the ratio $F_\star/F_0$, where $F_\star$ is the observed spectrum of a star seen through an interstellar cloud and $F_0$ is the spectrum  the star would have in the absence of intervening interstellar material, rather than in the magnitude form  $-2.5\log (F_\star/F_0)$ inherited from the traditional use of magnitudes\footnote{Both representations are  legitimate. Besides its historical origin, the $-2.5\log (F_\star/F_0)$ form provides direct access to $\tau_\lambda$, commonly used in interstellar dust modeling, whereas $F_\star/F_0$ corresponds more closely to the original data. Visually, the two representations differ markedly because the magnitude scale inverts and distorts  the original data. For clarity, I will refer to extinction curves in the traditional sense as mag-extinction curves. Hence, an ''exponential extinction curve'' corresponds to a ''linear mag-extinction curve''.}.
I will also distinguish extinction curves from extinction laws.
Although the two notions are generally conflated, observed extinction curves are not necessarily true physical extinction laws.
As noted by \citet[][CCM89 in the following]{CCM89}, the near-infrared continuum of reddened stars may be affected by thermal emission from nearby heated dust, introducing a nebular component into the infrared part of the extinction curves. 
Extinction laws instead represent the actual wavelength-dependent physical laws that determine the fraction of starlight absorbed or scattered by interstellar material along the line of sight.
For instance,  extinction curves of  the form $e^{-a/\lambda^p}$ ($p\simeq 1$, $a\propto E(B-V)$, see Section~\ref{expec})  introduced above are frequently observed in nature and  formalize the extinction laws of dust size distributions, set by the index $p$ and regardless of specific chemical composition  \citep[][]{vdh}.

The first part of this paper, Sections~\ref{expec}-\ref{ecdis}, reviews the essential properties of the two types of interstellar extinction curves that are currently observed: exponential extinction laws (Section~\ref{expec})  and Galactic-like extinction curves with a bump  (Section~\ref{galec}).
The relationship between the ultraviolet segment of extinction curves and the ultraviolet extension of their near-infrared-optical exponential extinction law  (see Sections~\ref{opgal}-\ref{ecgal})  leads to a classification of  all extinction curves, inside and outside the Galaxy, into one of three categories (Section~\ref{ecc}), as illustrated in Figure~\ref{fig:fig1}.
Section~\ref{bump}  examines the observable parameters commonly used to quantify extinction curves, $E(B-V)$,  total-to-selective extinction parameter $R_V=A_V/E(B-V)$, extinction coefficient at the bump position $A_{bump}$, and index $p$.
This Section questions their mutual relationships, as well as their connections with other observational tracers of interstellar clouds, including diffuse interstellar bands (DIBs), unidentified infrared bands (UIBs), metallicity, and carbon richness.
Completing this review, Section~\ref{ecdis} studies the relative distributions of Galactic bump-like and bump-free extinction curves throughout the universe.

A second part of the paper (Sections~\ref{centrism}-\ref{dis}) addresses the physical interpretation of extinction curves.
It is paradoxical that Galactic-like extinction curves, which are almost exclusively confined to Galactic sightlines and lack any known physical analogue, have been enshrined as the canonical reference for interstellar extinction, while the quasi-universal, bump-free, and physically consistent SMC-like extinction curves are  relegated to the status of anomalies (Section~\ref{centrism}).
This galactocentric perspective mirrors the former geocentric worldview and obscures a simpler interpretation of interstellar extinction observations that can be anticipated from Figure~\ref{fig:fig1}.
The gap, shown in grey in the figure, between observed ultraviolet extinction curves and the exponential extinction laws defined by their optical portion can be accounted for by forward-scattered starlight, implying that the extinction law underlying an ultraviolet Galactic extinction curve remains its exponential optical law (Section~\ref{explaw}).
This interpretation is in better agreement with the full set of observational constraints imposed on interstellar extinction curves, as established in the first part of this paper, than the prevailing assumption that the observed curves are extinction laws themselves.
It is, furthermore, the only interpretation consistent with the observed two-parameter dependence of ultraviolet interstellar extinction curves, another constraint that any viable explanation of interstellar extinction must meet (\citealt{param}; Sections~\ref{fits}-\ref{dparam}).

The discussion closing this paper  (Section~\ref{dis}) argues that, unless  the interstellar medium behaves in ways that contradict basic physical principles, 
the only sound explanation for the full range of extinction observations is the presence of a component of coherently forward-scattered starlight in the spectra of reddened stars.
The major implications are that Galactic-like ultraviolet extinction curves should not be conflated with extinction laws, and that interstellar extinction proper follows a single,  universal normalized  extinction law extending from the infrared to the far ultraviolet, with $R_V$ and $p$ of order  $R_V^0\simeq2.7$ and $p_0\simeq 1.4$.
This hypothesis can account for the peculiarities of Galactic ultraviolet extinction curves while also restoring simplicity and physical consistency to the interpretation of interstellar extinction data.

\section{Exponential-like extinction curves  extending over the optical spectrum} \label{expec}
Extinction curves  of the form $e^{-a/\lambda^p}$, which decrease exponentially over  at least the optical  range, obey the following relationships between the exponential index $p$ and $R_V$:
\begin{eqnarray}
R_V &=&\frac{1}{1.25^p-1}  \label{eq:rvp} \\
p&=&4.48\ln\left(1+\frac{1}{R_V}\right)  \label{eq:p}
\end{eqnarray}
Here, $A_V$ ($-2.5log_{10}(e^{a/\lambda_V^p}$)) and $A_B$ are defined using the Johnson $V$ and  $B$ bands, centered respectively at $\lambda_B = 0.44$~$\mu$m ($x_B=1/\lambda_B=2.27$~$\mu$m$^{-1}$) and $\lambda_V = 0.55$~$\mu$m ($x_V=1/\lambda_V=1.82$~$\mu$m$^{-1}$) such that $\lambda_B/\lambda_V \simeq 1.25$.

Equations~\ref{eq:rvp} and \ref{eq:p} show that, for extinction curves decreasing exponentially (linear mag-extinction curves) over the whole optical/near-ultraviolet domain (e.g. extinction curves with no 2200~\AA\ bump), a one-to-one relationship can be established between index $p$ and $R_V$, allowing $p$ to be derived from the determination of $R_V$.
For reference, $p$-values of approximately 2, 1.8, 1.6, 1.4, 1.2, 1,  and 0.8 will be associated with $R_V$-values near 1.78, 2.02, 2.33, 2.73, 3.26, 4.00, 5.12 respectively  \citep[see also Figure~2 and Table~1 in][]{rv2}, and reciprocally.

Still using the definitions of $A_V$ and $A_B$, one also finds
\begin{eqnarray}
a &=&E(B-V)\left(\lambda_B^{-p}- \lambda_V^{-p}\right)^{-1} \label{eq:a} \\
&=& \left(2.27^{p}- 1.82^{p}\right)^{-1}E(B-V), \label{eq:an}
\end{eqnarray}
wavelengths being taken in $\mu$m.
Hence, $a=2.22E(B-V)$ and $a=1.2E(B-V)$ for respectively $p=1$ and $p=1.4$.

It follows that  exponential extinction laws extending over the $[V, \,B]$ segment can be expressed in different ways
\begin{equation}
e^{-a/\lambda^p}=e^{-0.92A_\lambda}=  e^{-0.92R_V\left(\frac{\lambda_V}{\lambda}\right)^pE(B-V)}= e^{-\frac{a_p}{\lambda^p}E(B-V)} \label{eq:exp}
\end{equation}
with
\begin{equation}
a_p=0.92\lambda_V^pR_V= \frac{1}{\lambda_B^{-p}- \lambda_V^{-p}}. \label{eq:ap}
\end{equation}

Exponential-like extinction curves are physically meaningful over  their domain of validity.
They arise from dust media with a power-law grain size distribution,  $dN\propto a^{-v}da$ with  $v=p+3$ \citep[][and references therein]{vdh,shaw73}.
In diffuse media, such as terrestrial and planetary atmospheres or volcanic dust, such extinction laws are commonly observed, with $p\sim 1.4$ and the associated size distribution ($v\sim 4.3$) thought to be
 shaped by turbulence-driven particle growth and fragmentation, regardless of dust composition \citep[e.g.,][]{russell96,eck99, clancy03,tomasko08}. 
Beyond the Solar System,  whether spanning the entire near-infrared-to-far-ultraviolet spectrum or  valid only redward of the 2200~\AA\ bump (see Section~\ref{ecgal}),  they are also commonly observed and referred to as “SMC-type” curves, as they were initially characterized along sightlines to stars in the Magellanic Clouds (Section~\ref{ecdis}), and are particularly prevalent in the SMC ($R_V$ in the range $2.7-2.8$; \citealt{bouchet85,gordon03}).

\section{Galactic Extinction Curves} \label{galec}
\subsection{Galactic Near-Infrared and Optical Extinction Curves} \label{opgal}
The first decades (1930–1960) of interstellar extinction studies were limited to ground-based near-infrared and optical observations and established that extinction in the visible wavelength range is remarkably uniform throughout the Galaxy \citep{stebbins43}. 
Extinction  was found to depend solely on the amount of interstellar matter along the line of sight, as characterized by  the reddening $E(B-V)$.

From 1~$\mu$m to $\sim4500$~\AA, optical extinction decreases exponentially with inverse wavelength (see, for instance,  Figure~1 in \citealt{stebbins39}, Figure~2 in \citealt{ardeberg82}; Figure~14 in  \citealt[][]{nandy64}).
This dependence extends almost up to the $B$ band, with no appreciable deviation of individual sightlines from the mean  (see Nandy's Figure~15).
Today, it can readily be verified spectroscopically that the optical spectra of stars of the same spectral type differ exactly by a multiplicative exponential factor in $1/\lambda$,  implying an interstellar dust extinction law of the form $e^{-a/\lambda^p}$ with $p$ close to 1.

Beyond 1~$\mu$m,  however, near-infrared photometry in the  atmospheric transmission bands shows that Galactic infrared mag-extinction curves are not  linear in $1/\lambda$ but instead follow a  power law in $1/\lambda$ (\citealt{whitford48}; CCM89; \citealt{schultz75}).
Assuming continuity of extinction laws across the 1~$\mu$m boundary, infrared extinction observations, which are more sensitive to the value of $p$, correct the   linear dependence in $1/\lambda$ of the optical exponent.
The curvature of the exponent implies that  normalized near-infrared-optical extinction  follows  an exponential law $e^{-a_p/\lambda^p}$  with $p>1$ \citep{rv2}.
Although there is no reason for $p$ to differ from the Magellanic Clouds,  its exact value remains debated \citep{decleir22,butler24} and will be discussed in Section~\ref{p}.

\subsection{Galactic Ultraviolet Extinction Curves} \label{ecgal}

Ultraviolet observations of Galactic stars from above the atmosphere, in particular those of the International  Ultraviolet Explorer (IUE), confirmed the  indication from Figures 14 and 15 in \citet{nandy64} that toward near-ultraviolet wavelengths typical Galactic normalized mag-extinction curves systematically depart from their exponential optical extinction laws, as if extinction were less effective than predicted by a straightforward extrapolation of the law to shorter wavelengths.
These deviations usually become noticeable around 2.2-2.3~$\mu$m$^{-1}$ and are sensitive to direction.
Therefore, across the bridge $B$-band region marking the optical/near-ultraviolet junction, Galactic  extinction shifts from a one-parameter regime  ($E(B-V)$) to a multi-parameter regime.
The parameterization of this regime will be examined in Sections~\ref{fits}-\ref{dparam}.

\begin{figure*}
\resizebox{1.\columnwidth}{!}{\includegraphics{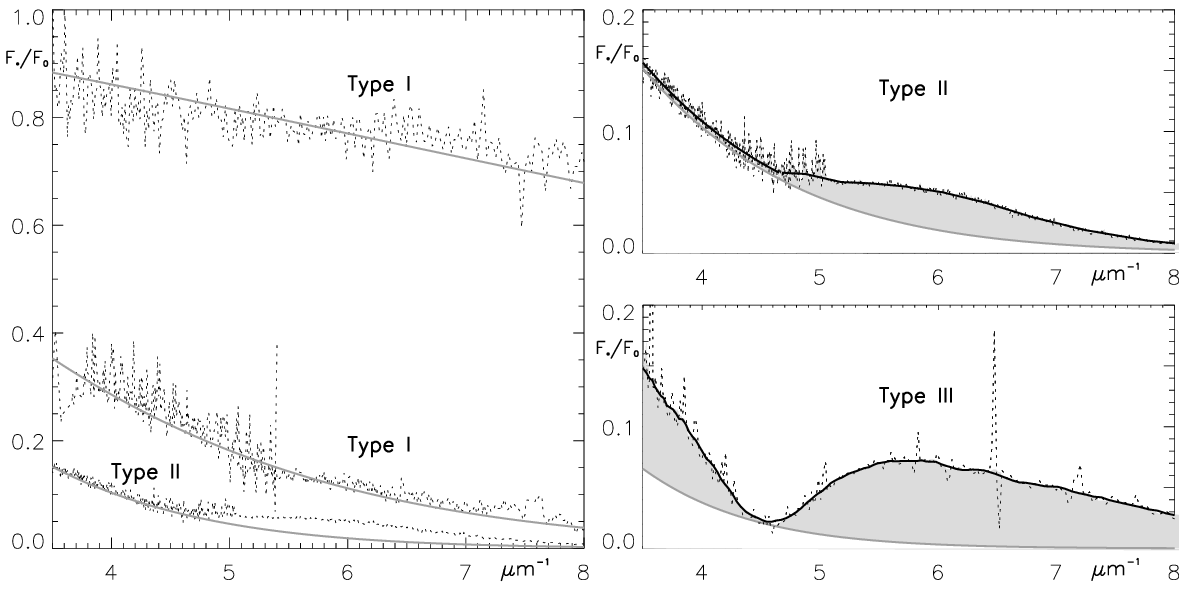}} 
\caption{Observed ultraviolet extinction curve types. 
All extinction curves fall into one of three categories according to the relationship they maintain in the ultraviolet with the exponential extinction law describing their optical part,  represented here by the gray exponential lines.
Type I (left panel) and Type II (left and upper right panels) curves remain closely associated with the optical exponential law, either throughout the ultraviolet spectrum (Type I) or down to the bump region only (Type II).
Type III curves (bottom right panel), the most common type in the Galaxy, clearly diverge from the optical exponential law, remaining above it except at the bump position (Section~\ref{bump}).
Gray shaded areas highlight the difference between observed extinction curves and their optical extinction laws and may reflect either a change in extinction law or an additional contribution of forward-scattered starlight (Section~\ref{explaw}).\\
High-redshift extinction curves are almost exclusively featureless Type~I curves.
In the Galaxy, Types I and II are observed only at extremely low reddening or when local extinction dominates. 
The examples shown here have reddening values $E(B-V)\sim 0.05$ and 0.17~mag  (Type~I), 0.3~mag (Type~II), and 0.45~mag (Type~III).
The exponential laws are $e^{-1.1E(B-V)/\lambda^{1.4}}$ (see Section~\ref{expec}).
} 
\label{fig:fig1}
\end{figure*}

Moving further into the ultraviolet, the curves increasingly deviate from  their underlying optical extinction laws, except in the bump region.
Section~\ref{bumpexp} shows that the bump only affects the part of the extinction curve that lies in excess of the exponential law, touching it at the bump position.
As illustrated by the lower right panel of Figure~\ref{fig:fig1}, an observed Galactic extinction curve therefore always remains above the exponential law defined by the optical part of its spectrum.

The near-ultraviolet  departure of extinction curves from the optical law implies that the observed value of $R_V$ ($=1/(A_B/A_V-1)$) must be larger than  the value inferred from the near-infrared-optical exponential law, since the curve deviates more from the exponential in the $B$ band than in the $V$ band.
Conversely, higher $R_V$ values indicate stronger departures  from the optical law in  the  near-ultraviolet (Figure~4 in CCM89).
Figures 1 and 2 in CCM89 also suggest that the shift between the near-infrared/optical and ultraviolet regimes is not abrupt and deserves closer examination. 
These figures indicate that deviations from the exponential law may already be present, albeit at a low level, at the longest wavelengths, and  then progressively increase with $1/\lambda$ until becoming clearly apparent near the optical-near-ultraviolet transition.

Galactic ultraviolet extinction curves that significantly  deviate in the near-ultraviolet from the exponential optical law are systematically characterized by the broad 2200~\AA\ absorption bump.
By contrast, a few  Galactic extinction curves lack the 2200~\AA\ bump or exhibit only a negligible feature.
For bump-less sightlines with very low reddening (typically a few hundredths of a magnitude), ultraviolet extinction curves of stars with the same spectral type differ by a simple exponential in $1/\lambda$ across the 3-8~$\mu$m$^{-1}$ wavelength range \citep[][]{vlc}.
This implies that, at very low reddening, the optical exponential extinction law $e^{-a/\lambda^p}$ extends into the ultraviolet.

\begin{figure}[h]
\resizebox{1\columnwidth}{!}{\includegraphics{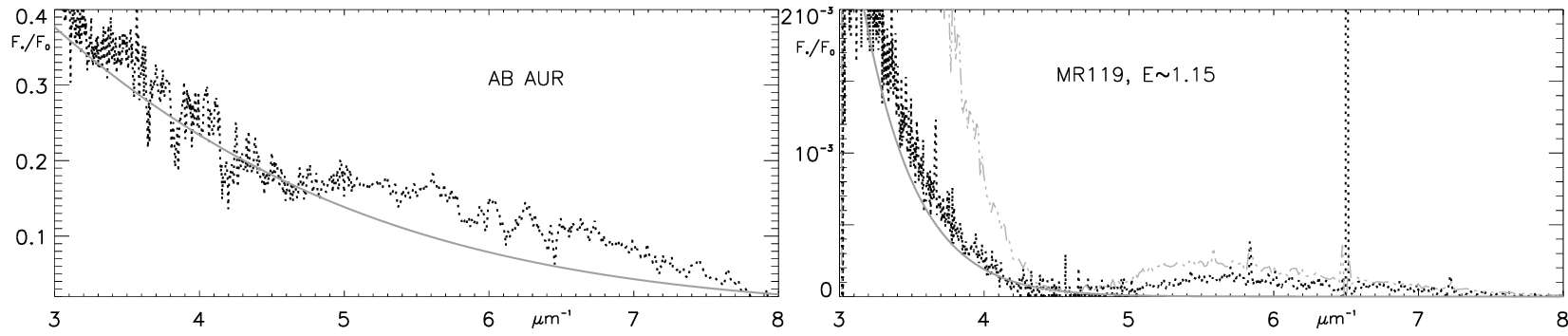}} 
\caption{Type~II extinction curves produced by local dust.\\
\emph{Left:} Ultraviolet extinction curves of the shell star AB~Aur (HD~31293, $E(B-V)\simeq 0.2$; \citealt{sitko81}).
With the exception of one case, all extinction curves derived by Sitko et al. remain close to the optical exponential law up to the bump, as illustrated here by the extinction curve toward AB Aur.
\\
\emph{Right:} Ultraviolet extinction curve of Wolf-Rayet star MR~119, affected by strong local extinction   \citep[$E(B-V)\simeq 1.15$; ][]{garmany84}.
Although Garmany et al. did not explicitly derive the extinction curve, they noted that applying a standard extinction correction to their  Wolf-Rayet sample would yield a far-ultraviolet thermal turnover redder than observed, implying that circumstellar ultraviolet extinction must be substantially stronger than predicted by standard models.
The extinction curve of MR~119 closely follows the gray exponential line corresponding to the exponential extinction law $e^{-1.1\times 1.15/\lambda^{1.4}}$ down to the bump region, and displays a much stronger pre-bump extinction than expected for a normal Galactic field star, such as the Type III star HD~229196  ($E(B-V)\simeq 1$; upper gray dotted curve).
} 
\label{fig:fig2}
\end{figure}

Similarly, stars embedded in  circumstellar dust or stars illuminating a nearby foreground nebula show no, or only a strongly reduced, 2200~\AA\ bump  \citep[][see Figure~\ref{fig:fig2}]{sitko81,garmany84}.
\citet{garmany84}  noted that the spectra of a sample of T-Tauri stars show no bump and cannot be explained unless their extinction is significantly stronger than that of typical field stars (right panel of Figure~\ref{fig:fig2}).
In all  cases where stars  are primarily obscured by nearby material, extinction curves remain close to the optical exponential laws well into the ultraviolet, at least up to the 2200~\AA\ bump region.
To my knowledge, these two situations, very low column densities and  sightlines containing a substantial proportion of local dust, account for all reported examples of small or vanishing  2200~\AA\ bumps.
Field stars located in the far background of interstellar clouds with sufficiently large column densities, by contrast, systematically exhibit a bump.
\subsection{Classification of Ultraviolet Extinction Curves Based on Their Deviation from the Optical Exponential Law } \label{ecc}
It follows from Sections~\ref{expec}-\ref{galec} that all stellar extinction curves, whether Galactic or extragalactic, can be categorized into three types based on how closely their ultraviolet segment aligns with the extension of the optical exponential extinction law into the ultraviolet.

Type I extinction curves follow the exponential law throughout the ultraviolet spectrum. 
This type largely dominates outside the Galaxy (see Section~\ref{ecdis}).
An extinction curve is classified as Type II if it remains close to the exponential law across the near-ultraviolet, down to the bump region. 
These curves are also frequent in the Magellanic Clouds \citep{mc}.
In the Galaxy, Type I and II  curves are found only along very low column density sightlines, or when part of the intervening interstellar matter lies close to the star (Section~\ref{ecgal} and Figure~\ref{fig:fig2}).
The Type III pattern corresponds to the classical Galactic extinction curves, which were the first to be observed and are rarely seen outside the Galaxy  (Section~\ref{ecdis}).
Type III curves deviate from the exponential law on both sides of the 2200~\AA\ bump.

In all cases where an extinction curve departs from the optical exponential  law, it remains above the exponential, touching it at the bump extremum (Figure~\ref{fig:fig1}; Sections~\ref{ecgal} and \ref{bumpexp}).
This implies that, for the same reddening,  a Type~I curve or the near-ultraviolet segment of a Type~II profile lies below a Type III curve, giving the appearance of stronger ultraviolet extinction, as noted by \citet{garmany84} and illustrated in Figures~\ref{fig:fig1} and \ref{fig:fig2}.

In the Galaxy and in the Magellanic Clouds, stellar extinction curves corresponding to a given  $E(B-V)$ may exhibit diverse shapes, yet they consistently fall into one of three categories: Type I, II, or III. 
\section{Observables of Interstellar Extinction: $E(B-V)$,  2200~\AA\ Bump Strength, $R_V$, and $p$; their Relationships and Dependencies} \label{bump}
The previous sections have introduced several observable variables that commonly appear in interstellar extinction studies: $E(B-V)$, $R_V$, $A_{bump}$, and the index $p$.
This section provides additional detail on these variables, their mutual relationships and their correlation with DIBs and UIBs, and their supposed dependence on local physical conditions (metallicity, carbon richness).
\subsection{$E(B-V)$} \label{ebv}
Reddening $E(B-V)$ is certainly the extinction parameter  most easily and directly accessible from observations.
It quantifies  extinction by interstellar dust and satisfies, with excellent precision,  the  relationship established by \citet{bohlin78}
\begin{equation}
N(H)+2N(H_2)\simeq 6\,10^{21}E(B-V),
\label{eq:boh}
\end{equation}
which has since been repeatedly verified \citep{chen15,liszt14,lacy17}.
This relation shows that $E(B-V)$ is a tracer of the total column density of interstellar matter on the line of sight.
It also implies that dust is intimately mixed with gas in the Galactic interstellar medium.

Reddening also enters the optical extinction law $e^{-a_pE(B-V)/\lambda^p}$, inasmuch as the $V$ and $B$ belong to the exponential regime (Section~\ref{galec}).
It follows from Section~\ref{ecgal} that, for Galactic directions with a 2200~\AA\ bump, the observed $E(B-V)$ is  slightly lower than the $E(B-V)$ of the  optical exponential law.
\subsection{The 2200~\AA\ Bump Relation to $E(B-V)$ and to the Optical Exponential Extinction Law} \label{bumpexp}
The 2200~\AA\ bump never appears in  exponential extinction curves.
The first deviations from the exponential law appear in the far-ultraviolet (Type~II curves), starting at the position of the bump.
When this departure extends only slightly into the near-UV, the bump still reaches its extremum on the exponential law.
This finding suggests that the bump only affects the part of an extinction curve above the optical exponential law.
Supporting this idea, published mean normalized Galactic mag-extinction curves show that  the extrapolation of the optical mag-exponential law into the UV passes close to the bump extremum \citep[see Figure~1 in][]{vlc}.

This result can be obtained analytically.
Early ultraviolet extinction studies by \citet[][]{bless72,savage75,nandy76} showed  that the average bump-height-to-$E(B-V)$ ratio, estimated in different regions of the sky,  remains constant throughout the Galaxy.
This ratio follows directly from  Equations 3a and 3b of CCM89 applied  to the bump extremum at $x=4.6$~$\mu$m$^{-1}$ 
\begin{equation}
A_{bump}\simeq 9.89E(B-V)
\label{eq:abCCM89}
\end{equation}
(in the CCM89 framework, $a(4.6)=-0.003\simeq0$).

This is essentially the value predicted by an exponential law $e^{-a/\lambda^p}$ around $p\simeq1.4$,
\begin{equation}
A_{bump}\simeq 10.12E(B-V),
\label{eq:aeexp}
\end{equation}
from Equations~\ref{eq:an}-\ref{eq:exp}.
As a function of $p$, the ratio $A_{bump}/E(B-V)$ reaches a minimum around $p\simeq1.3-1.4$ and increases  on either side. It equals 10.22 at $p=1$, and 10.88 and 11.50 at $p=1.8$ and $p=2.0$, respectively. 

Therefore, extinction curves that follow the CCM89 relation, that is 94~\% of Galactic directions according to \citet{valencic04}, remain just above their underlying exponential laws in the vicinity of the bump, and the two curves are essentially tangent at the bump position.
Type~II extinction curves obviously also retain the same property, as well as any combination of Type I, II, and CCM89 extinction curves on the same line of sight, thus presumably encompassing all observable extinction curves (see \citealt{mc} and Section~\ref{zfit}).
Physically, the bump removes almost exactly the portion of the extinction that  lies above the exponential law in Figure~\ref{fig:fig1}.
\subsection{$R_V$} \label{rv}
$R_V$ depends on the total extinction $A_V$ and is therefore the least straightforward parameter to determine.
It has most often been estimated from empirical relations involving near-infrared photometry and is consequently subject to both observational and methodological uncertainties.
These uncertainties explain why, although the determination of $R_V$ was instrumental in the construction of the CCM89 parameterization of Galactic extinction curves, markedly  improved fits within the CCM89 framework (or similar formalisms, see Section~\ref{fits}) are obtained when  $R_V$ is treated as a free parameter rather than as a fixed observable.

As noted at the end of Section~\ref{ecgal}, $R_V$ indicates the departure of ultraviolet extinction curves from the exponential extinction law defined in the optical.
$R_V$ reflects the change of slope of extinction curves, that is, the amplitude of this deviation per unit reddening, whereas the total amplitude of the deviation is more directly captured by $E(B-V)$ (Section~\ref{bumpexp}).
Figure~7 of CCM89 also highlighted a correlation between the bump strength normalized by $A_V$, $A_{bump}/A_V$, and $R_V^{-1}=E(B-V)/A_V$.
This correlation has been presented as problematic in previous studies, from CCM89 to \citet{zhang25}, although it follows trivially from the $A_{bump}$-$E(B-V)$ correlation.

\subsection{Bump Correlation with DIBs} \label{bumpdib}
DIBs are known to arise from the edges of interstellar clouds \citep[the skin effect;][and reference therein]{herbig95}.
They are observed in very low reddening high latitude directions \citep[down to a few hundredths of a magnitude,][]{baron15} that are essentially devoid of molecular gas and must therefore originate in HI media.
Like the bump, DIBs tend to weaken in circumstellar environments \citep{krelowski97,dibs}.
The 2200~\AA\ bump is also observed at low reddening, suggesting that it too forms in HI clouds. 
Since DIBs and bumps appear together in stellar spectra, their strength can be directly compared.
Studies by  \citet{nandy75,cox07} demonstrated that bump and DIB strengths are indeed  correlated. 
\subsection{The Bump/DIBs Versus  UIBs Paradox} \label{bumpuib}
The relation between UIBs and the 2200~\AA\ bump or DIBs is ambiguous.

In the Galaxy, UIBs can be observed wherever HI interstellar matter is exposed to ultraviolet radiation, from photodissociation fronts (PDRs) at the edge of interstellar clouds, with densities of the order of $10^4$~cm$^{-3}$ and illuminated by strong radiation fields \citep{habart24},  to   the diffuse interstellar medium of the Galactic disk or  high latitude cirrus with densities well below 1 H-atom/cm$^3$ \citep[][] {onaka96, lemke98, kahanpaa03}.
UIBs are also noticeably absent from ionized and molecular gas \citep{habart24}.
Furthermore, when  \citet{massa22} compared   UIBs from  fields centered on 16 background stars to the bump in the stars' spectra, they found a correlation between the bump and some UIBs.
\citet{zhang25} and \citet{gordon24} reached similar conclusions for several fields in the Galaxy and for the Magellanic Clouds.
UIBs therefore share the same HI environments as bumps and DIBs.

However, UIBs are barely observed in the same directions as bumps and DIBs.
UIBs do not appear in emission together with bumps and DIBs in the extinction spectra of background field stars.
They are particularly strong  in reflection and planetary nebulae, such as the Red Rectangle, NGC7027, NGC7023, NGC2023, IC63, etc..., where they accompany extended red emission (ERE) and the Red Rectangle bands (RRBs).
Toward the central stars, however,   DIBs and bumps are either absent or much weaker than expected (Sections~\ref{ecgal} and \ref{bumpdib}), and are generally attributed to foreground dust \citep{dibs}.

Likewise, galaxies generally show UIBs but almost never exhibit the 2200~\AA\ bump (see Section~\ref{ecdis}).
For instance, in the \citet{draine07} study of the infrared emission of 65 nearby galaxies, most  display Milky Way-like infrared spectra. 
Several of these galaxies have IUE observations, and all those with  ultraviolet spectra above the noise level,
 Mrk33 ($q_{PAH}=1.9$)\footnote{Here $q_{PAH}$, which in the authors' model measures the fraction of dust mass in the form of PAHs,  is essentially an indicator of the galaxies' UIB strength. $q_{PAH}$ spans the range from 0.47\% to 4.6\%. Regardless of its value, however, no 2200~\AA\ bump is detected in these galaxies.}, NGC4569  (4), NGC1097 (3.1), NGC 1705 (0.6), NGC3351/M95 (3.2), NGC4736 (4.1), NGC5194 (4.5), NGC7793 (3.6), lack a 2200~\AA\ bump. 

Hence, although UIBs, DIBs, and bumps all appear to arise in HI regions, they do not coexist: DIBs and the bump are weak or absent along sightlines where UIBs are strong, and vice versa.
This apparent contradiction between observations along specific sightlines and over larger areas admits two possible explanations. 
It may indicate that UIBs and bumps (or DIBs) have distinct carriers that respond differently to local physical conditions, such as variations in the radiation field.
Alternatively, the discrepancy may result from the geometry of observation: UIBs are detected in reflection and most easily near bright stars, whereas DIBs and bumps are seen in absorption toward distant background stars.

The importance of  geometry in interpreting observations of interstellar matter was already emphasized by  \citet{dibs}, and this is the more likely explanation.
In IC63, for instance, ERE and UIBs are seen throughout the nebula, whereas DIBs appear only along the line of sight to a background star \citep[Section~7 in][]{dibs}.
This finding supports the idea that the detection of UIBs versus that of the bump and DIBs depends more on viewing geometry than on intrinsic differences between their carriers.

\subsection{Other Bump Dependencies} \label{bumpred}
The tight correlation between the  2200~\AA\ bump and $E(B-V)$ in the Galaxy precludes any dependence of the bump on local interstellar physical conditions such as metallicity, carbon abundance, or stellar winds, contrary to suggestions by \citet{rocca81, gordon98, junkkarinen04, draine07, ma18, ormerod25}, among others.
If the bump were controlled by metallicity, the Galactic bump-$E(B-V)$ correlation would not hold. 
 \citet{jura77} provided a counter-example arguing against a metallicity-bump connection in quasars.
 Likewise, large surveys of galaxies spanning wide metallicity ranges consistently report the absence of the 2200 Å bump  \citep{kinney93, calzetti94,gordon97,valencic03,noll05}.
 
The bump region hosts a multitude of metal absorption lines, which can easily lead to spurious associations between the bump and metallicity.
At high redshift, these lines compromise continuum determination  (\citealt[][]{calzetti94}), thereby casting doubt on recent claims of high-$z$ bump detections (Section~\ref{ecdis}).
More than two decades ago  \citet{pitman00} broadly rejected claims of high-redshift bump detections  on the ground that apparent bumps in QSO spectra may arise from intrinsic QSO features produced by blends of Fe II emission multiplets on either side of the bump region.

The association of the bump with carbonaceous material, whether graphite grains or PAHs \citep{draine89, joblin92, steglich11}, remains equally  speculative and lacks both observational support and laboratory confirmation.
Section~\ref{bumpuib} further showed that the PAHs responsible for the UIBs cannot be the carriers of the 2200 Å bump, while \citet[][and references therein]{uibs} established that UIBs are unrelated to carbon-rich environments.
\subsection{The exponent index $p$ of Galactic Near-infrared/Optical Exponential Extinction Laws} \label{p}
In the extragalactic universe, extinction curves are predominantly exponential laws of the SMC type (Section~\ref{ecdis}), characterized by a well-established value $R_V^0$ in the range 2.7-2.8, corresponding to an  exponential index $p_0\simeq1.4$ (Section~\ref{expec}).
In the Milky Way, the optical index $p$  can best be determined along sightlines where optical extinction laws extend into the ultraviolet,  i.e. those without a 2200~\AA\ bump, such as  HD62542 or HD210121.
For these stars   \citet{valencic04} found $R_V=2.82\pm0.24$ and $2.42\pm 0.29$ respectively, consistent with the SMC value and an optical-law index $p$ close to 1.4 \citep[see also][]{valencic03}. 

Exponential extinction laws should also correspond to a Galactic minimum in $R_V$ (Section~\ref{ecgal}).
The study by  \citet{turner14} was motivated by the search for this minimum   and concluded that it lies  close to 2.8.
More recently,  the  all-sky $R_V$ map derived from optical data for over a hundred million stars  by \citet{zhang25sc} yields a minimum $R_V$ of about $2$, which would imply $p\sim 1.8$.
However, uncertainties are large, and the minimum of averaged $R_V$ values points instead to a minimum close to 2.7, as illustrated by the averaged $R_V$ values of  Figure~4 in \citet{zhang25sc} and Figure~3 in \citet{zhang25}.

Finally, Galactic $R_V$ can  be estimated from near-infrared  data if Galactic infrared extinction follows the same normalized law as in the optical, which is expected in the absence of any physical reason for a discontinuity at 1~$\mu$m.
All near-infrared studies, which are almost exclusively ground-based and photometric, consistently show that near-infrared extinction is uniform across the Galaxy, yet published values of $p$ vary widely \citep[see][and references therein]{rv2,maiz20, hensley23}.
This scatter is largely driven by the low level of infrared extinction and by calibration issues affecting ground-based photometry (CCM89, \citealt{rv2}).
Nevertheless, the comparative study of several photometric datasets carried out by D.~Turner and myself led us to conclude that $ 1<p<1.6$, with a clear preference for $p\simeq1.4$ \citep[see Figure~3 in][]{rv2}.
By contrast, more recent near-infrared photometric studies  tend to favor larger values, up to  $p\simeq 2$ \citep{maiz20,butler24}.
Such values are difficult to reconcile with a smooth optical–infrared transition, as illustrated in Figure~3 of  \citet[][]{rv2}, and would push $R_V$ towards the extreme low end of its observed range ($p>1.6$ implies $R_V< 2.3$; Section~\ref{expec}). 
To date, the only spectroscopic  investigation of near-infrared extinction, by  \citet{decleir22}, finds values of $p$ between 1.36 and 2.08 for a sample of 15 stars.

Overall, while a unified exponential  law extending from the near-infrared into the optical with $p=p_0\simeq 1.4$ appears plausible, the precise connection between optical and near-infrared extinction remains to be clarified.
Resolving this issue will require spectral studies covering extinction curves from the near-infrared to at least the near-ultraviolet \citep[as tentatively approached using photometry in][]{rv2}, as well as targeted investigations of  low-$R_V$ sightlines.
Neither has yet  been undertaken.

\section{Spatial Distribution of Bump and Bump-Free Extinction Curves in the universe} \label{ecdis}
The first large compilation of normalized mag-ultraviolet extinction curves  ($E(\lambda-V)/E(B-V)$ versus $1/\lambda$) for Galactic stars, published by \citet[][their Figures~5-6]{bless72},  revealed  a systematic  Type III pattern of Galactic extinction curves.
Shortly after, however, deviations from the Galactic extinction behavior were identified in extragalactic sources as an absence of the 2200~\AA\ bump in quasar and Magellanic Cloud spectra  \citep{mckee74,borgman75, jura77}. 
\citet[][their  Figure~1]{borgman75} demonstrated that the average mag-extinction curve for stars in the Large Magellanic Cloud (LMC) followed a nearly linear trend from the near-infrared to the far-ultraviolet, thereby establishing the existence of Type I extinction curves, linear in magnitude and devoid of a bump.
A similar conclusion was reached by  \citet[][see their Figure~1]{prevot84} for the Small Magellanic Cloud, where the average normalized mag-extinction curve was also found to be linear and bump-free.
Star-by-star analysis of extinction curves in the Magellanic Clouds has now established that most LMC and SMC stars are bump-free and have a Type~I or Type~II extinction curve \citep[][]{gordon03,mc}.
Section~\ref{ecgal} recalled that Type I-II extinction curves are also present within the Galaxy, but only under two specific circumstances: when the column density is very low or along sightlines toward stars obscured by dust in their immediate vicinity.

Outside the Galaxy, the early finding by  \citet{mckee74} reporting the absence of the 2200 Å bump in 38 quasar spectra was, and continues to be confirmed by larger surveys.
\citet[][]{york06b} found no bump in 809 lightly reddened quasars ($E(B-V)<0.1$~mag),  while  \citet[][see their Figure~14]{glikman12}  reached the same conclusion for 120 more heavily reddened  ($0.1<E(B-V)<1.5$) ones.
 In both samples,  extinction was best described by SMC-like curves.
Similar results hold for starburst galaxies  \citep[][Figures~20-21]{calzetti94}, $\gamma$-ray bursts \citep[GRBs,][]{heintz17,zafar18}, as well as for the few hundreds galaxies, including Galactic-like spirals, reviewed by \citet{kinney93} and \citet{zeimann15}.
Although the determination of extinction curves in redshifted quasars and galaxies is hampered by the absence of reliable dust-free references for comparison, their ultraviolet spectra are exponential-like and consistent with Type I extinction laws \citep{gordon98}.

Until the early 2000s, observations of Galactic-like extinction curves (exhibiting the 2200~\AA\ bump) outside the Milky Way were limited to galaxies within the Local Group, most notably the Magellanic Clouds \citep{rocca81, lequeux82, rosa94}.
No large survey of quasars or high-redshift galaxies conducted prior to 2000 reported convincing detections of the bump, and the few early claims of Galactic-type extinction curves at high redshift were critically rebutted by \citet{pitman00}.
Moreover, the true origin of the observed Local Group bumps remains ambiguous, as these lines of sight cannot clearly disentangle Galactic foreground extinction from extinction intrinsic to the external galaxies  \citep{gordon03}.
For instance, the Magellanic Clouds are seen through Galactic cirrus with average reddening toward the LMC  ($E(B-V)\sim0.08$)  roughly twice as high as that toward the SMC \citep[see  discussion in][]{mc}.
This could explain the higher proportion of Galactic-like extinction curves in the LMC.

The first robust case of Galactic-like extinction at cosmological distance was reported by  \citet[][]{junkkarinen04}.
They identified both a prominent 2200 Å bump and DIB $\lambda$4430 in the spectrum of QSO AO 0235+164  ($z_{em}=0.94$), attributed to an intervening absorber at $z_{abs} = 0.52$ approximately halfway between the observer and the quasar. 
This finding was independently confirmed through the identification of several strong DIBs at the expected rest-frame wavelengths of the absorber  \citep{york06a, lawton06, ellison08}\footnote{No DIBs have been detected in other QSO/DLA systems, except in quasar SDSS J001342-002412 ($z_{em}=1.644$)  associated with a Ca II absorber at $z_{abs}=0.1556$ \citep{ellison08,chang25}.}. 
A few more bump detections in  Damped Lyman-$\alpha$ (DLA) systems have been reported \citep{srianand08, wang04,zhou10, jiang11, wang12} with the absorber well in front of the background QSO.
But, to date, the Junkkarinen et al. DLA remains the only known case at high redshift where both bump and  DIBs have been reliably detected.

\citet[][]{jiang11}  identified only 39 DLAs with a possible 2200~\AA\ bump out of 2951 quasar sightlines.
In a more recent and much larger survey, \citet{zhang25s} reported 843 cases (0.15\%)  with weak 2200~\AA\ features among $5.56\,10^6$ quasar spectra ($0.7<z<2.4$) from the Sloan Digital Sky Survey. 
These detections exhibit narrow profiles and displaced peak positions.
In the authors’ own words, they require further confirmation.
None of the 239 high-redshift galaxies ($1.90 <z <2.35$)  examined by \citet{zeimann15} exhibited a bump.
Among 514 $\gamma$-ray bursts (GRBs) analyzed by \citet{greiner24}, only six showed signs of a 2200~\AA\ bump, and just one of those appeared consistently across all epochs following the afterglow.  
However, this latter detection remains uncertain owing to the absence of any of the strong DIBs expected to appear redshifted in the near-infrared in spectra obtained by \citet[][]{zafar18apj} with the European Southern Observatory’s X-shooter spectrograph.

A few recent claims of 2200~\AA\ bump detections at high redshift, based on limited samples of sightlines, are  doubtful, as the reported features depart too strongly from the defining properties of the Galactic bump, which are also observed at high redshift in QSO AO 0235+164  (see Figure~1 in \citealt{junkkarinen04}).
In  \citet[][]{noll09}, the conclusion that about 30\% of 78 luminous galaxies at $1 < z < 2.5$ exhibit a bump is undermined by the strongly reduced width of the feature, more than 60\% narrower than in the Galaxy.
\citet[][]{witstok23} and \citet[][]{ormerod25} each report a single high-redshift galaxy with a bump whose central wavelength is shifted by more than 80~\AA\ relative to the Galactic value, a displacement incompatible with the properties of Galactic bumps.
Doubts may also arise regarding the weak, marginal, and metallicity-correlated depressions in the 2200~\AA\ region reported by  \citet{shivaei22} (see the red and yellow spectra in their Figure~4), derived from stacked    $z\simeq 2$ galaxy spectra $\left(E(B-V)\sim 0.1-0.2\right)$.
These depressions are more naturally explained by a blurring effect from the dense metal absorption lines in this wavelength range during stacking (genuine 2200~\AA\ bumps are unrelated to metallicity; see Section~\ref{bumpred}), and no individual spectra claimed to exhibit 2200~\AA\  bumps are presented.

These tentative high-redshift bump detections are generally driven by attempts to link the 2200~\AA\ feature to UIBs and/or metallicity and ultimately to lend support to PAH-based models of interstellar dust.
In practice, however, since the properties of the 2200~\AA\ bump are grounded in Galactic extinction studies,  high-$z$ bump claims will remain speculative unless they fulfill the bump shape properties and the concomitant presence of DIBs, known to systematically accompany bona fide ultraviolet bumps, can be demonstrated in the same spectra.
To date, this concerns a single direction.

In any event, 
 at the scale of the universe, observations separate Galactic sightlines largely dominated by Type~III extinction curves with a bump from extragalactic ones that, with few exceptions, lack a 2200~\AA\ bump and are most generally Type~I curves, presumably similar to the SMC extinction.

\section{Galactocentrism} \label{centrism}
Since the beginning of interstellar extinction studies, the terms extinction curve and extinction law have been used interchangeably. 
In this respect observations reveal two primary types of extinction laws in the universe: a Galactic extinction law, marked by the prominent  2200~\AA\ bump and dominant within the Milky Way, and a quasi-universal, SMC-like exponential extinction law, which prevails throughout the rest of the universe (and within the Solar System).
Type III extinction curves dominate Galactic sightlines but remain the exception rather than the rule across the observable universe (Section~\ref{ecdis}).  
Yet they have been canonized as the standard of interstellar extinction  and  serve as the foundational template for dust models.
Meanwhile Type I  featureless extinction curves, which account for the overwhelming majority of extragalactic observations have consistently been labeled  "anomalous" or ''peculiar'' \citep{borgman75,desert90,gordon03}.

This galactocentric perspective has distorted the interpretation of cosmic extinction.
The absence of the 2200~\AA\ bump in galaxies and QSOs has not been assessed on its own terms, but rather interpreted as a deviation from the Galactic norm, explained away through a patchwork of speculative, galaxy-specific factors such as metallicity, carbon abundance, gas-to-dust ratio, stellar mass, or turbulence  \citep[][among others]{bless72araa, borgman75, koornneef84, prevot84, clayton00aspc, shivaei22, witstok23, ormerod25, zhang25s}.
None of these factors has ever been shown to influence the 2200~\AA\ bump (Section~\ref{bumpred}) or to account for the differences between Type III and Type I/II curves observed within the Milky Way itself.

As a result, exponential extinction curves have been overlooked for what they are: the extinction signature of a power-law size distribution of scatterers, independent of any specific dust composition (Section~\ref{expec}). 
This, together with the underappreciated connection between Galactic extinction curves and exponential laws (Sections~\ref{ecc} and \ref{bumpexp}) illustrated in Figure~\ref{fig:fig1}, calls for a reversal of perspective on interstellar extinction.
Type I extinction curves should be taken as the baseline behavior across the universe and the specific conditions under which the comparatively rare Type III curves arise should be investigated.

\section{Figure 1 and the  Interpretation of  Galactic Extinction Curves } \label{explaw}
With respect to the exponential laws defined by their optical part, sightlines with  Types~II and III curves  exhibit  an excess of light reaching the detector.
The  gap (shown in light-grey in the right panels of Figure~\ref{fig:fig1}) between the exponential baseline and the observed  extinction curve may reflect a change in the extinction law, as traditionally assumed.
Alternatively, it could  be filled by an additional contribution of forward-scattered starlight superimposed on the extinguished  stellar flux.
This possibility would also account for the fact that these extinction curves do not correspond to the extinction law of any known particle or combination of particles, either on Earth or elsewhere in the universe.

The possibility that scattered light contributes to the observed spectrum of reddened stars was first raised by \citet{snow74}.
It was dismissed on the grounds that ultraviolet observations from different telescopes showed no beam-size dependence of the spectra. 
However, this argument does not hold under the specific conditions of near-complete forward scattering, which is coherent over extremely small angles, well below the angular resolution of any telescope.
For a star at infinity and an interstellar cloud located at a distance $L$ from Earth, the coherence  domain, defined by the first Fresnel zone\footnote{The first Fresnel zone is the region around the line of sight to the star within which all forward-scattered waves from atoms interfere constructively at the observer’s location. For a star at infinity and a scattering screen at distance L, the zone has a radius $\sqrt{\lambda L}$ and an area $\pi \lambda L$. In Fresnel optics, roughly half of this area contributes effectively to the forward-scattered beam.} at the cloud position, is subtended by an angle $\theta\sim\sqrt{\lambda/L}$, yielding values on the order of $5\,10^{-13}$~rad, or $10^{-7}$~arc-second, for $1/\lambda\sim 7$~$\mu$m$^{-1}$  (1430~\AA) and a cloud 100~pc away. 
This angle is far below any current telescope resolution. 
Yet, the Fresnel zone itself covers a large physical  area, roughly 1500~km in diameter in this configuration. 
Coherent scattering from this region would scale as the square of the number of hydrogen atoms within the zone, rather than linearly as in the case of incoherent scattering, leading to a significant amplification of the scattered signal.

This  component of forward-scattered starlight in the spectrum of reddened stars must remain weak when the optical depth is low, whether at large  wavelengths (in the infrared and, to a lesser extent, in the optical) or along low column density sightlines, or when the first Fresnel zone shrinks because of the proximity between a star and the scattering medium.
All these dependencies  are consistent with Galactic observations (Section~\ref{ecgal}). 
Since the 2200~\AA\ bump is tied to the scattered component (Section~\ref{bumpexp}), it follows that interstellar matter intrinsic to a light source should not produce a bump in the source’s spectrum.
Consequently, galaxies and QSOs should not display a 2200~\AA\ bump in their ultraviolet spectra unless a foreground absorber intervenes at a distance comparable to the absorber-earth distance, as in the only known case,  the BL Lac AO0235+164, where bump and DIBs were detected (Section~\ref{ecdis}).


\section{Empirical Fits of Extinction Curves and Their Implications} \label{fits}
\subsection{FM Multi-Parameters Against CCM89 One-Parameter Fits of Galactic Normalized Mag-Extinction Curves} \label{2fits}
The characteristic shape of Galactic-like Type III extinction curves lends itself well to analytical representation. 
Yet no physically grounded derivation has so far convincingly reproduced their far-ultraviolet behavior.
Since the 1990s, two polynomial parameterizations in $x=1/\lambda$, both lacking physical justification, have dominated the representation of interstellar extinction curves:  the multi-parameter FM fits \citep{fitz88,fitz05,fitz07}, and the one-parameter  CCM89 expression of Galactic normalized mag-extinction curves.
Despite their empirical nature, the FM and CCM89 fits have become standard tools for representing extinction curves
and are widely used to draw inferences about variations in interstellar dust properties. 

Although often used side by side, the FM and CCM89 fits are fundamentally antagonistic.
FM fits can achieve excellent representations of Galactic and SMC-like extinction curves, but only at the cost of a large number of parameters (up to ten, depending on the version, see \citealt{param}). 
In contrast, CCM89 concluded that variations among Galactic-like extinction curves could be captured by reddening $E(B-V)$ and the single additional parameter $R_V$ alone. 

A ten-parameter fit and a two-parameter fit are clearly not the same and reflect incompatible assumptions about the nature of Galactic interstellar matter.
The FM approach rests on the premise that infrared/optical extinction, the 2200~\AA\ bump, and far-ultraviolet extinction arise from distinct populations of particles.
For instance, based on the FM parameters, \citet[][]{jenniskens93} argued that four distinct, independent classes of particles are needed to explain extinction curves.
In clear opposition,  CCM89 conclude that modification of the extinction curve at one wavelength induces a  change across the entire curve, thereby contradicting the idea that multiple independent components contribute separately to interstellar extinction.

This tension between the FM and CCM89 approaches crystallizes a debate that had remained latent since the earliest studies of ultraviolet interstellar extinction. 
The local spatial variability of extinction curves was already emphasized in the large early survey by  \citet{bless72}, suggesting that multi-component grain models might be required to account for such diversity.
However, Figures~3 to 5 in  \citet{nandy76} showed linear correlations between $E(\lambda-V)$ and $E(B-V)$ at three ultraviolet wavelengths  (2740, 2190, 1392~\AA) for a sample of  four hundred reddened stars.
The authors interpreted these trends as evidence of a relatively uniform ultraviolet extinction law across the Galaxy, consistent with Nandy's earlier results in the optical.

Interpreted analytically and extrapolated to all ultraviolet wavelengths, the correlations found by Nandy et al. between $E(\lambda-V)$ and $E(B-V)$ suggest the existence of two functions, $a_0(\lambda)$ and $b_0(\lambda)$, such that
\begin{eqnarray}
E(\lambda-V)&\simeq &b_0(\lambda)E(B-V)+ a_0(\lambda) \nonumber \\ 
&\simeq & b_0(\lambda)E(B-V) \label{eq:nandy}
\end{eqnarray}
since $A_\lambda=0$ when there is no reddening ($E(B-V)=0$).
Introducing successively $\lambda_B$ and $\lambda_V$ in this equation, it also follows that $b_0(\lambda_B)=1$ and $b_0(\lambda_V)=0$.

Equation~\ref{eq:nandy} can be rewritten
\begin{equation}
\frac{A_\lambda}{A_V}\simeq \frac{b_0(\lambda)}{R_V}+1,
\label{eq:nandy2}
\end{equation}
If $R_V$ remains close to 3, as \citet{nandy76} might have assumed, ultraviolet extinction would appear uniform through the Galaxy, as \citet{nandy64} found for the optical domain.
This uniformity, however,  breaks down when $R_V$ varies.
Figures~3 to 5 in  \citet{nandy76} actually suggest that Galactic ultraviolet extinction is  not governed solely  by the amount of reddening (quantified by either $A_V$ or $E(B-V)$), but requires a single additional adjustable variable ($R_V$ in Equation~\ref{eq:nandy2}).

Soon after this breakthrough, \citet{mrn} showed that Galactic extinction curves could be reproduced through the introduction of several grain size distributions (implying several extinction laws and  multiple free parameters), and \citet{koornneef78} questioned the spatial homogeneity of Galactic extinction inferred by Nandy et al., though he did  not dispute  the $E(\lambda-V)$ versus $E(B-V)$ relationships. 
\citet{greenberg83} normalized several  Galactic mag-extinction curves to the same near-ultraviolet wavelength and showed that their  far-ultraviolet rises differed markedly \citep[their Figure~4; see also Figure~2 in][]{param}. 
They interpreted these differences  as evidence for three distinct interstellar extinction laws, each dominating a specific portion of extinction curves--the optical range,  the 2200~\AA\ bump, and the far-ultraviolet rise--and linked  to an independently varying particle population.
This view corroborated the \citet{mrn} model and naturally privileged a multi-parameter description of interstellar extinction, later formalized in the FM parametrization with its many adjustable coefficients.

Yet  \citet{greenberg83} made no reference to Nandy et al.'s $E(\lambda-V)$ versus $E(B-V)$ relationships.
If these relations hold, when considered together with the tight correlation between bump strength and $E(B-V)$ (Section~\ref{bump}), they invalidate  the multi-components interpretation of interstellar extinction proposed by Greenberg \& Chlewicki, without contradicting their observations.
As Equation~\ref{eq:nandy2} shows,  extinction curves normalized to a given wavelength may diverge in shape while still depending on a single parameter beyond reddening.
Figure 4 in  \citet{greenberg83}  therefore demonstrates neither the complete independence of the curve’s segments nor the existence of multiple carriers and  does not justify the  many free parameters introduced in the FM fit.

Six years after the \citet{greenberg83} paper,  and independently of the \citet{nandy76} results, CCM89 revisited the problem of interstellar extinction dependency by searching for correlations between extinction at different wavelengths.
They found that a modification at any wavelength of a Galactic normalized mag-extinction curve affects the entire curve in a systematic way that depends solely on $R_V$.
Echoing Equation~\ref{eq:nandy}, they approximated normalized extinction curves  $A_\lambda/A_V$ with an empirical relationship of the form\footnote{As for Equation~\ref{eq:nandy}, Equation~\ref{eq:CCM89}  should normally imply $b(x_V)=0$ and $a(x_V)=a(x_B)=b(x_B)=1$.
Neither the functions $a(x)$ and $b(x)$ given by CCM89, nor their subsequent revisions, from \citet{valencic04} to \citet{gordon24}, satisfy these relations.
In practice, considering that the uncertainty on the determination of $R_V$ is relatively large, $R_V$ should be considered  an adjustable parameter in the CCM89 fit \citep{valencic04,mc,param}.}
\begin{equation}
\frac{A_\lambda}{A_V}\simeq a(x) + \frac{b(x)}{R_V},
\label{eq:CCM89}
\end{equation}
where $x=1/\lambda$.

The relationship was verified with surprising accuracy for several dozen stars, for which CCM89 carried out  as precise a determination of  $R_V$  as was possible.
\citet{valencic04}  applied their own determination of $a(x)$ and $b(x)$ in Equation~\ref{eq:CCM89}  to  a sample of 417 Galactic sightlines with reddening larger than 0.2~mag.
Assuming that $R_V$ was an arbitrary parameter, not necessarily the one derived from near-infrared observations (see Section~\ref{rv}), they found that 97\% of these directions followed the relationship, with only four stars--Type~III HD204827, and Type~II HD62542,  HD29647, and HD210121--showing significant deviations.
All four sightlines are characterized by a strong component of local reddening.

These results indicate that Galactic Type~III extinction curves  generally conform to the CCM89 fit.
Nandy et al.’s relationships (Equation~\ref{eq:nandy}) have also been confirmed from the infrared to the far-ultraviolet  with remarkable precision by \citet[][their Figure~3]{bondar06} for a sample of 22 B1V stars, and more recently by \citet[][his Figures~9 and 10]{param}, over a larger dataset comprising more than 550 sightlines. 
\subsection{A notable improvement over CCM89 fits: the \citet{mc} fit of extinction curves} \label{zfit}
The CCM89 fit was developed specifically for sightlines toward Galactic field stars and is, with few exceptions, restricted to that context. 
In each direction,  the strength of the 2200~\AA\ bump is determined by the reddening, while the near-ultraviolet change of curvature  is governed by  $R_V$ (see Section~\ref{bump}).
As such, the CCM89 formalism is inherently incapable of handling Type I and Type II extinction curves, where the absence of a bump  implies $E(B-V)=0$.
Another key limitation of CCM89 fits is that they are usually proposed in a normalized form, whether by $A_V$ or $E(B-V)$,  which implicitly assumes a homogeneous dust medium along the line of sight. 
Some directions, however,  may combine  Type I/II and Type III extinctions, such as when both local and foreground dust media are  present along a star’s line of sight, a situation the CCM89 framework  does not account for.

The Magellanic Clouds provide a unique environment where individual stars display a balanced mix of Type I, II, and III extinction curves. 
In \citet{mc}, I showed that Magellanic extinction curves with identical bump strength but differing ultraviolet slopes superimpose well once corrected by an exponential law accounting for the difference in stellar reddening.
The CCM89 fit was then improved by dividing a star line of sight into a CCM89-like component with reddening $\Delta_1$, fixed by the size of the 2200~\AA\ bump and the corresponding $R_V$-dependent CCM89 relationship, and by an exponential-like extinction for the remaining $\Delta_2=E(B-V)-\Delta_1$ reddening.
The resulting fit matches most Magellanic extinction curves and also improves the Galactic fits obtained by \citet{valencic04}.
Cases such as HD204827  that resisted the \citet{valencic04}  fit can now be successfully modeled \citep[Figure~9 in][]{mc}.

The \citet{mc} fit of extinction curves  still depends on reddening and parameter $R_V$ alone, and takes into account the size of the bump. 
It  fits virtually all extinction curves with a bump with excellent accuracy \citep[see][]{mc} and enlarges its domain of application to Type~I  exponential extinction curves.
It follows that almost all Galactic and Magellanic extinction curves can be represented by the same two-parameter expression, which should have considerable importance for comparing  their interstellar medium from extinction observations.   

Type~II curves, such as those toward HD62542,  HD29647, HD210121 are the only ones that cannot be reproduced by this combined CCM89 and exponential fit.
They lack a significant bump to support a CCM89 component, and their exponential decline does not extend beyond the bump region. 
This limitation, however, can be considered a direct consequence of the empirical nature of the fit.
\section{What can the physical origin of the free parameter of interstellar extinction be?} \label{dparam}
Reddening alone characterizes Type~I extinction curves, whereas a second parameter is required to describe Type III curves (Section~\ref{zfit}).
Although $E(B-V)$ clearly traces the amount of material along the line of sight,  the physical origin of the second parameter remains to be identified.
Within the Milky Way, the only physical factor shown to affect normalized extinction curves is the star’s proximity to foreground dust.
The proximity of interstellar matter alone appears sufficient to transform standard Galactic curves into bump-less, SMC-like profiles.
Apart from reddening, no other physical parameter has been observed to control the presence or absence of the 2200~\AA\ bump (Section~\ref{bump}), and therefore the type of extinction curve.
Consequently, if, as follows from Section~\ref{fits}, Galactic extinction curves depend on both the column of interstellar matter and a second parameter, that parameter is most plausibly the star–intervening-dust distance.

The influence of distance on the spectra of reddened Galactic stars has never been considered critical in interstellar dust modeling, which links modifications of extinction curves to variations of grain size distributions and composition.
Interstellar dust models are tied to the tripartite structure of extinction curves described in \citet{greenberg83}, which requires two to four interstellar dust components, and, whatever   their nature, on the order of a dozen free parameters to reproduce the curves, regardless of the model considered \citep{desert90, jenniskens93, weingartner01, hensley23}.
That these parameters should align systematically with both the value of $R_V$ and the distance between the star and the occulting material is unrealistic. 

While distance has little meaning as a fundamental parameter in conventional interstellar dust models, it naturally emerges as the intrinsic  parameter of coherent forward scattering (Section~\ref{explaw}) for stars observed behind Galactic HI clouds.
Beyond the Milky Way, the relevant free parameter should be the size of the first Fresnel zone, which depends on the distances between the light source, the intervening interstellar matter, and the observer.
For a fixed source–observer distance, the efficiency of coherent forward scattering is maximized when the scattering medium lies halfway between them. 
For a given distance between the interstellar matter and the observer, maximum efficiency is reached for sources located at distances of the order of, or larger than, this value.
These conclusions are consistent with observations both in the Galaxy (Section~\ref{ecgal}) and at high redshift  (Section~\ref{ecdis}).

\section{Discussion}  \label{dis}
One of the major difficulties in the study of interstellar extinction today lies in the accumulation of speculative ideas that, over the past decades, have gradually hardened into accepted conventions.
In particular, the frequently asserted link between the 2200~\AA\ bump and metal-rich or carbon-rich environments, or the claim that bumps and UIBs could arise from absorption and emission by the same particles (e.g. PAHs), are both manifestly unsupported by observations  (Sections~\ref{bumpuib}-\ref{bumpred}).

These widely circulated ideas stem largely from modeling expectations and from the 1984 proposal--never demonstrated--that UIBs originate from PAH-like molecules \citep{leger84,allamandola85}.
Forty years later, however, no one can cite a single PAH contributing to the 2200~\AA\ absorption.
Likewise, the notion that fullerenes such as C$^+_{60}$, C$_{80}$, C$_{540}$ \citep{iglesias07,cordiner19}, or polyacenes like  C$_{48}$H$_{26}$ \citep[][]{omont19}, invoked as potential DIB carriers, populate high-latitude cirrus with $E(B-V)$ values of only a few $10^{-2}$~mag \citep{baron15} can be seriously questioned. 

In this paper, I have revisited the key observed properties of interstellar extinction curves, as they were already known or accessible by the 1980s, with particular emphasis on how ultraviolet Galactic extinction curves relate to their underlying near-infrared–optical extinction laws (Section~\ref{opgal}).
In contrast to optical observations, ultraviolet extinction observations oppose two main and fundamentally different behaviors of interstellar extinction curves  (Section~\ref{ecdis}).
Featureless, exponential Type I curves  (Section~\ref{expec}) extend the optical law far into the ultraviolet spectrum and still depend on reddening only.
The more complex Type III curves, dominant along Galactic sightlines, are characterized by the broad 2200~\AA\ bump  and significant spatial variability, even when normalized to unit reddening (Section~\ref{ecgal}).
The spatial dependence of Type III curves, however, can be fully accounted for by introducing an additional parameter, usually  associated with $R_V$,  alongside $E(B-V)$ (Section~\ref{fits}).
Altogether, the full range of Type I exponential and Type III Galactic extinction curves can be described within a single analytical formulation that accounts for mixed extinction components along the same line of sight and depends only on reddening and the additional parameter (Section~\ref{zfit}).

Between Types I and III, some intermediate extinction curves (Type II), also lacking a bump, remain close to the optical law down to the bump region and exhibit a deficit of extinction only in the far-ultraviolet (Section~\ref{ecgal}). 
In the Galaxy, bump-free, SMC-like Type I and Type II extinction curves occur only under two circumstances: at very low column densities, or when interstellar matter lies close to the observed star. 

The graphical representation of Type I, II, and III ultraviolet extinction curves in Figure~\ref{fig:fig1} provides a clear overview of their defining properties and mutual relationships.
For a given reddening, Type II and III curves always lie above the underlying extinction law (Type I) set by their exponential optical extinction (Figure~\ref{fig:fig1}; Section~\ref{ecc}).
These curves further touch the extinction law at the bump extremum, demonstrating that the bump removes only the part of the extinction curve that rises above the exponential law   (Section~\ref{bumpexp}).
Figure~\ref{fig:fig1} also highlights two possible interpretations of Types II and III curves. 
Either the curves themselves represent the true extinction laws along the sightlines, as traditionally assumed, or the underlying Type I exponential laws remain the genuine extinction laws into the ultraviolet, with the observed deficit in extinction (light grey area in the plots) resulting from an additional contribution of coherent forward-scattered starlight from the medium along the line of sight  (Section~\ref{explaw}).

Over the past fifty years, considerable efforts have been directed toward providing physical consistency to Type III extinction curves, assumed to be extinction laws,  even though, unlike Type I exponential curves, they have no physical counterpart on Earth or in the Solar System and, on a cosmic scale, appear more as local anomalies than as widespread laws (Sections~\ref{ecdis}-\ref{centrism}).
From the MRN size distributions \citep{mrn} to the most  recent \citet{hensley23} model, interstellar dust models have succeeded in reproducing Type III extinction curves but have failed to capture the most fundamental properties of interstellar extinction.
No model can reduce its number of free parameters below a dozen without internal inconsistencies, which is clearly at odds with two-parameter fits of extinction curves (Section~\ref{zfit}; \citealt{param}).
In the Galaxy, models have failed to address the weakening or disappearance of the bump, that is, the systematic transformation of extinction curves from Type III to Type I or II in low column-density or locally extinguished sightlines (Section~\ref{ecgal}).

Moreover, the potential carriers of either the bump or the far-ultraviolet rise are still uncertain and the subject of much controversy \citep{kwok23}.
That such a wide variety of suggested exotic molecules\footnote{In addition to fullerenes and polyacenes, PAHs are further believed to exist in multiple forms: neutral or ionized; dehydrogenated, hydrogenated, or super-hydrogenated; deuterated or super-deuterated; nitrogen- or oxygen-substituted (PANHs, O-PAHs);  as clusters or fragments; and so on. } can survive and even proliferate in near-vacuum high-latitude cirrus and show no spectroscopic signature in molecular clouds is certainly puzzling.
\citet{zhang25} proposed that their formation might be triggered in situ. 
Yet, the probability of such chemistry occurring in media with densities between 0.01 and 100 H-atom/cm$^3$ \citep{kahanpaa03} must be virtually zero.
The suggestion by \citet{witstok23}  that PAHs appeared in galaxies at $z\sim 7$, presumably formed in the carbonaceous ejecta of Wolf–Rayet stars or supernovae, is no more credible:  Wolf–Rayet stars generally lack the 2200~\AA\ bump (Figure~\ref{fig:fig2}; \citealt[][]{garmany84}); UIBs are not correlated with carbon-rich environments   \citep[and particularly carbon-rich supernovae,][]{uibs}; and recent galaxies, including Milky Way–type spirals, also lack the bump  (Section~\ref{ecdis}). 

The assumption that extinction curves are extinction laws is therefore in deep conflict with elementary logic. 
In the alternative view, where scattered starlight generally contaminates the observed spectra of Galactic stars, scattering is forward-directed and must be coherent; its strength then depends on column density  and on the size of the coherence area, which is set by the star–cloud distance (Sections~\ref{explaw} and \ref{dparam}).
Local extinction, in particular, should contribute little, if anything, to the scattering.
The scattering component of extinction curves diminishes at low optical depth and with decreasing star–cloud distance, corresponding to the two situations in which weaker or vanishing bumps are effectively observed (Section~\ref{ecgal}). 
In such cases, optical extinction laws extend down to the bump region or even throughout the ultraviolet spectrum.

Extinction curves thus appear to depend on two parameters: the  reddening $E(B-V)$ and the fraction of scattered starlight in the spectrum, reflected by the divergence of extinction curves from the optical extinction laws, and potentially indicated by the value of $R_V$  (Section~\ref{ecgal}).
$E(B-V)$ clearly traces the total amount of matter along the line of sight.
In the Galaxy, extinction curves with observed values of $R_V$  larger than its minimum value $R^0_V$ (corresponding to the absence of scattered starlight; Sections~\ref{ecgal} and \ref{dparam}) must include a proportion of scattered light and therefore depend on the distance between the obscuring interstellar matter and the star.

This distance, through its effect on the size of the first Fresnel zone,  is the additional quantity that  a physically meaningful  derivation of extinction curve profiles will need to include.
In the most general case, including observations across the wider universe, the relevant physical parameter becomes the relative distance between the light source, the intervening interstellar matter, and the observer.
This naturally explains why the only firm detections of 2200~\AA\ bumps at high redshift, those for which DIBs are also detected, occur in DLAs where the bump is associated with galaxies located well in front of the background quasar  (Section~\ref{ecdis}).
It also explains the significant improvement in extinction-curve  fitting achieved by the approach described in Section~\ref{zfit}, which, as a first approximation, separates the interstellar matter between a star and an observer into local and distant components. 

These results extend a series of conclusions published since 2020 to the continuum of interstellar extinction.
\citet{param} emphasized the need to reduce interstellar dust extinction to a two-parameter description, while \citet{uibs,ere,dibs} showed that the major interstellar emission and absorption features\footnote{UIBs, extended red emission (ERE), Red Rectangle bands (RRBs), and blue luminescence (BL) in emission; DIBs and the 2200~\AA\ bump in absorption.} superimposed on the infrared-optical part of the continuum can be related to the processing of ultraviolet radiation by atomic hydrogen through Raman or inverse Raman scattering, depending on the observational geometry.
The spectrum of nebular matter illuminated by strong ultraviolet-dominated radiation fields exhibits a characteristic hydrogen Raman emission spectrum, whereas observations of distant ultraviolet sources intercepted by intervening interstellar matter reproduce, on cosmic scales, the inverse Raman scattering experiment of \citet{jones64} and its corresponding absorption spectrum.

Altogether, this series of papers recenters the physics of the interstellar medium on hydrogen and on the main atomic constituents of HI interstellar clouds, effectively excluding the presence of complex molecules such as PAHs or their analogues in these environments.
It also introduces the geometry of the observation as a parameter in its own right that must be explicitly taken into account.
These results  further suggest that the interaction of atomic hydrogen with starlight in space still requires investigation and could ultimately provide a laboratory-tested explanation for several long-standing problems raised by observations of the interstellar medium.
Potential applications include corrections to interstellar extinction, distance estimates, and the determination of the atomic composition of interstellar H I media.
The present study suggests that a single extinction law, close to $e^{-1.2E(B-V)/\lambda^{1.4}}$, may describe the continuum of interstellar dust extinction across the universe.

Acknowledgements: This work greatly benefited from the contributions of Jane F. Bestor and ChatGPT in the editing, organization, and documentation of the manuscript.
\newpage

\bibliographystyle{model3-num-names}
{}
     \end{document}